\documentclass[12pt,3p]{elsarticle}

\usepackage{natbib}
\usepackage{lineno,hyperref}
\usepackage{amsmath,amssymb,mathrsfs}
\usepackage{graphicx,subfigure}

\usepackage{multirow}

\modulolinenumbers[5]

\journal{Journal of \LaTeX\ Templates}

\begin{document}

\begin{frontmatter}

\title{Effect of observational holes in fractal analysis of galaxy survey masks}

\author[label1]{J. E. Garc\'ia-Farieta\corref{cor1}}
\cortext[cor1]{I am corresponding author}
\ead{joegarciafa@unal.edu.co}
\author[label1]{R. A. Casas-Miranda}
\ead{racasasm@unal.edu.co}
\address[label1]{Departamento de F\'isica, Universidad Nacional de Colombia - Sede Bogot\'a, Av. Cra 30 No 45-03, Bogot\'a, Colombia}

\begin{abstract}
Cosmological observations reveal that the Universe has a hierarchy of galaxy clustering with a transition to homogeneity on large scales according to the $\Lambda$CDM model. On the other hand some observational estimates suggest a multifractal behavior where galactic clustering is based on generalization of the correlation dimension. From this point of view, we study the influence of veto areas on fractal measurements in masks of galaxy surveys. Particularly we investigate if these holes can produce fractal behaviors or modify the scale of cosmic homogeneity. From the footprint of the Baryon Oscillation Spectroscopic Survey (BOSS) data release (DR12), we build a homogeneous sample following the radial selection function for $73,412$ points limited to the redshift range $0.002<z<0.2$. Different percentages of observational holes were created cumulatively in right ascension and declination on the sample. For the synthetic sample and for a real sample of galaxies we determined the fractal dimension $D_q(r)$ in the range $2\leq q\leq6$ using the sliding window technique to characterize the spatial point distribution. Our results show that generalized dimension varies with the scale, for low scales there are a fractal behavior with fluctuations for all hole percentages studied and for larger scales than $113~Mpc/h$ the statistical homogeneity is achieved in concordance with other analysis. We find that observational holes cause a shift in the homogeneity scale $r_H$, in particular for all synthetic samples with percentages of holes between $0-10\%$ the homogeneity scale is reached at $(83\pm1)$Mpc/h while the fractal dimension changes as $2.83^{\pm0.09}\leq D_q\leq2.855^{\pm0.09}$. For synthetic samples with percentages of holes greater than $10\%$, we find that the value of $r_H$ increases proportionally. Consequently future results about homogeneity scale based in fractal analyses must be corrected by observational holes and regions of incompleteness in the geometry of the galaxy catalogue if the size of the veto mask is significant.
\end{abstract}

\begin{keyword}
large-scale structure\sep galaxy clustering\sep observational holes\sep multifractal analysis
\MSC[2010] 85A40\sep 85A35\sep  28A80
\end{keyword}

\end{frontmatter}


\section{Introduction}
The large-scale structure of the Universe has been extensively studied since the publication of the first galaxy surveys. Some analysis with high-redshift samples show the existence of fractal correlations \citep{celerier2005fractal,labini1998scale, joyce1999fractal, Martinez445}, where the large network of matter filaments that connect galaxies is characterized by fractal structures \citep{martinez2010statistics, JonesMartinez1988,  Pietronero2000PhyA, Pietronero2001ctap}. In a cosmological context, fractals were introduced into a model by Mandelbrot as a hypothesis to solve the Olbers paradox \citep{martinez1990universe}. According to this idea, if the set of stars forms a structure with self-similarity properties such as a Cantor dust the paradox is solved, for even in a mathematically infinite universe, the sky can have dark regions. We currently know that at small scales, $r\lesssim10^{-1}~Mpc$, the galaxy distribution exhibits a fractal behavior \citep{peebles1980large,peebles1993principles,mandelbrot1982the}, although some observational estimates suggest that at larger scales the structure is more complex and can be characterized as a multifractal object, requiring a generalisation of the fractal dimension in order to explain this distribution on the same scale range \citep{conde2015fractal}, in this sense the fractal concept allows interpretation of the hierarchical clustering in terms of self-similarity properties or an invariant scale of the galaxy distribution \citep{chaconstudy}.\\

Fractal analysis is an useful mathematical tool that quantifies galactic clustering using data from galaxy surveys by calculating quantities such as the fractal dimension, making it possible to establish relationships between these values and other statistical descriptors. The possible cosmological implications of fractal analysis of the galaxy distribution are discussed in detail by \citet{baryshev1998fractal} and \citet{martinez1991fractal}. Because cosmic clustering developed under the influence of gravity alone, there is no physical motivation to consider smaller scales related to the processes of galaxy formation. In this context, the multifractal measures are described as a natural scaling result of the matter interactions in the Universe. A formal and general proof of the fractal nature of the matter distribution on large scales requires additional observations and some solutions of the Einstein field equations taking into account the conditional cosmological principle \citep{martinez1990universe}. An additional motivation to study the matter distribution on large scales from a multifractal viewpoint is the transition to a homogeneity scale, which can be defined according to \citet{yadav2010fractal} as the value of $r$ above which fractal dimension $D_q$ of the distribution is equal to the dimension of the physical space in which the points are distributed, i.e., $D=3$; this implies that if the distribution is finite in size and weakly clustered, such as that in the first galaxy catalogues, it is difficult to achieve this equality. For a non-integer dimension the galaxy distribution is in a state of transition to homogeneity \citep{grujic2006fractal, grujic2009fractal}.\\

The clustering description under the fractal hypothesis has encouraged the development of several theoretical models. One of the first models was proposed by \citet{mandelbrot1982the} using L\'evy flights to simulate the distribution of galaxy clusters regardless of underlying physical phenomena. Similarly, \citet{peebles1989fractal} developed the basis for a statistical description of structures in the Universe, using correlation functions of up to four points for the first galaxy catalogues, he showed that the fractal dimension depends on the size of the distribution. Hence, this quantity changes from a pure fractal with dimension $D=1.26$ for $r<15.33~Mpc$ to structures consistent with the standard cosmological principle on larger scales \citep{peebles1989fractal}. Some observations have shown that the galaxy distribution grows in proportion to $r$ raised to an exponent related to the correlation dimension. For the galaxy distribution on a significant range of scales this dimension is approximately 2 \citep{martinez2010statistics}, a very different value from 3, which is what would be observed for a homogeneous distribution because the number of galaxies grows in proportion to the volume of a sphere as expected according to the standard cosmological model given the homogeneity and isotropy conditions.\\

Recent studies have also reported fractal and multifractal behavior of the spatial distribution of galaxies and dark matter halos using galaxy surveys in which large-scale fractal structures are evident \citep{pino1995evidence}; in particular, see \citet{zheng1988fractal} for \emph{CfA survey data}, \citet{xia1992fractal} for the \emph{IRAS survey}, and \citet{seshadri1999multi} for \emph{the Campanas redshift survey}. \citet{chacon2012millennium} performed a multifractal analysis of dark matter halos from the \emph{Millennium} simulation \citep{springel2005simulations}, they found a transition to homogeneity between $100$ and $120~Mpc/h$ in strong agreement with the LCDM model. In contrast, an analysis with volume-limited samples from SDSS-DR7 \citep{chaconstudy, chacon2016multi, munoz2012galaxy} reported fractal behavior at large scales until $165~Mpc/h$, where $D_q$ is always below the homogeneity limit $D=3$ for all values of the structure parameter. This result is consistent with \citet{joyce1999fractal, labini2009absence}, who found hierarchical patterns with self-similarity properties and a fractal dimension smaller than the dimension of physical space at scales greater than $100~Mpc/h$. For fractal analysis with WiggleZ, \citet{scrimgeour2012wigglez} reports a transition to homogeneity at $r_H=71\pm8~Mpc/h$ with $z\leq0.2$; this indicates that the galaxy distribution does not behave as a fractal object. This result is also consistent with \citet{hogg2005cosmic}, \citet{yadav2005testing} and \citet{sarkar2009scale}, who report a transition to homogeneity at $\sim70~Mpc/h$. \citet{wu1999large} and \citet{yadav2005testing} are in agreement with this value for the homogeneity scale, however, their studies show that at smaller scales the galactic cluster has fractal properties with dimension $D\approx1.2-2.2$. In all cases the effects of the geometry of the surveys must be taken into account according to \citep{yadav2010fractal,Pan01032002}. In particular, the fractal calculations may be affected by the presence of holes and borders in the catalogues that are inherent to the process of observation using astronomical instruments.\\ 

In this paper we studied the multifractal behavior of a spatial distribution of points using synthetic samples built from the Sloan Digital Sky Survey III (SDSS-III) - BOSS  footprint \texttt{boss\_survey.ply}, including a random distribution spatially uncorrelated of observational holes related with the veto masks in right ascension and declination, in order to compare our results we also analyzed a real galaxy sample. We determined the fractal dimension in the range $2\leq q\leq6$, the multifractal dimension spectrum, and the homogeneity scale using the sliding window technique for each sample. In section 2 we define the main concepts of fractal formalism such as the fractal dimension, multifractal description of galaxy clustering and generalized fractal dimension. In section 3 we review the origin of observational holes and footprints of BOSS masks. Moreover, we present the construction of synthetic samples limited in redshift, including the distribution of holes. Then in section 4 we present and discuss the results obtained about the application of our method to the constructed samples with the BOSS footprint. Finally the conclusions are given in section 6.

\section{Fractal and multifractal formalism}
\subsection{Concepts of fractal dimension}\label{sec:fractalconcept}
The concept of dimension can be associated with the number of degrees of freedom or the minimum number of coordinates to specify any point within a distribution of points in a metric space. Topologically, the dimension indicates how much space a set occupies near each of its points \citep{falconer2004fractal}. The most intuitive definition of dimension is the \emph{topological dimension} $D_T$, introduced by Poincare and generalised by Lebesgue. In this definition, given a set of topological space $\mathbb{X}\in\mathbb{R}^n$, the dimension is the minimum value of $n$ for which every open cover admits a locally finite open refinement the order which does not exceed $n+1$. If there is not a minimum value of $n$, we say that the set is infinite-dimensional. 
In Euclidean space, $\mathbb{R}^n$, $D_T$ takes integer values in $(0,~n)$ \citep{mandelbrot1982the}.\\

When the sets describe irregular shapes the concept of dimension in terms of the number of coordinates is insufficient to describe them. This fact has motivated the introduction of new concepts beyond the classical geometry \citep{mandelbrot1982the}, therefore, fractal geometry was developed. Thus, it is possible to give a different concept of \emph{dimension}, for instance, the \emph{self-similar dimension} can be explained by further fragmentation of an object or set, and the ratio of the number of identical parts where each part is scaled down by the ratio $r$. For any set $X~\in~R^n$ that supports division into a finite number of subsets $N(k)$, where all of them are consistent with each other by translations and rotations, and it is a reduced copy of the initial set by a factor $r=1/k$. The self-similar dimension of $X$ is defined as the unique value $D$ satisfying the equation $N(k)=k^D$ \citep{mandelbrot1982the}, i.e.,
\begin{equation}\label{eq:DimensionFractal1}
D=\frac{\log N}{\log (k)}.
\end{equation}
Here $D$ is not necessarily an integer number. In some cases it may be an integer and match the topological dimension. When the object cannot be subdivided into exact copies of itself we can use the \emph{box-counting dimension}, in this case a set $\mathscr{A}$ is covered by a grid or regular boxes with side $\delta>0$, all equal to each other, then the number of boxes $N(\delta)$ needed to cover the figure is determined. This process is very natural for a computer, and it is not necessary that the figure be self-similar. Hausdorff and Besicovitch \citep{mandelbrot1982the} proposed a more general definition of \emph{dimension} that considers fractional values and can be defined for any set of points. Mandelbrot defined a fractal object as a set with Hausdorff dimension $D_H$ strictly exceeding its topological dimension. Thus sets with non-integer Hausdorff dimension are fractals. In practice the Hausdorff dimension is not always easy to calculate  \citep{falconer2004fractal}. In this case, following the idea of the \emph{self-similar fractal dimension}, the \emph {mass-radius fractal dimension} $D_m$ is defined by a power law; this dimension is the measure of the total mass contained in a sphere of radius $R$ whose center is a point of the set, and the mass contained as a function of the radial size is determined as
\begin{equation}\label{eq:DimensionMasa1}
M(R)=FR^{D_m}, 
\end{equation}
where the factor $F$ is a function that may be different for fractals with identical dimension, and the density number of galaxies decreases as $n(R)\approx R^{D_m-d}$ for a set in $\mathbb{R}^d$ \citep{martinez2010statistics}. 

\subsection{Multifractals and generalised fractal dimension}
For point distributions from galaxy catalogues the analysis is more complex than for modeled distributions because fluctuations associated with the intrinsic characteristics of the mass distribution appear, so we need to use more general geometric estimators \citep{william2000distribution}. Multifractals provide the most detailed description possible of the fractal properties of a distribution of points. Given a distribution in which each region exhibits a fractal behavior, but the dimension changes from one place to another, it is possible to establish correlations between these dimensions and to do a complete analysis from a higher level \citep{blumenfeld1997levy}.\\

For physical properties that depend on the scale the fractal behavior at small scales reported by \citet{peebles1989fractal}, \citet{bagla2008fractal}, and \citet{martinez1990universe} can be extended to a matter distribution at large scales using the multifractal formalism. For each center of the point distribution,  the number of particles $n_i(r)$ contained within a sphere of radius $r$ measured from the position of the $i$th particle is given by
\begin{equation}\label{eq:ni}
n_i(r)=\sum_{j=1}^N \Theta(r-|\mathbf{x_i}-\mathbf{x_j}|),
\end{equation}
where the sum is over all particles in the sample, and $N$ is the total number of particles. The coordinates of each particle in the three-dimensional space are denoted as $\mathbf{x_j}$, $j\neq i$, and $\Theta$ is the Heaviside function. The number of particles $n_i(r)$ around each galaxy taken as the center, with coordinates $\mathbf{x_i}$, is determined by counting the particles around the center that are located within a comoving sphere of radius $r$ \citep{celerier2005fractal}, that is, a sphere expanding with the Hubble flow, where the distance between two points remains fixed as the universe expands.\\

The correlation dimension is defined similarly to the mass-radius fractal dimension \citep{seshadri2005fractal}. To characterize the distribution, we must have all the information about the statistical moments in order to define the generalized dimension. The generalized correlation integral $C_q(r)$ is defined as
\begin{equation}\label{eq:Cq}
C_q(r)=\frac{1}{NM}\sum_{i=1}^M[n_i(r)]^{q-1}, 
\end{equation}
where $q$ is called the structure parameter and corresponds to an arbitrary real number, and $M$ is the number of particles used as centers. According to \citet{murante1997density}, from the correlation integral it is possible to perform a power series expansion of $\log{(r)}$ [equation~(\ref{eq:ExpanLOG})] and thus to calculate directly the multifractal dimension $D_q$. It is sufficient to keep the first two terms on the right side of equation~(\ref{eq:ExpanLOG}), which is simplified so that a simple relation between the generalized correlation integral and generalized fractal dimension is obtained as in equation~(\ref{eq:Cqrq}) \citep{chaconstudy}.
\begin{equation}\label{eq:ExpanLOG}
\log[C_q^{1/(q-1)}]=D_q\log(r)+\log(F_q)+\mathcal{O}\left(\frac{1}{\log(r)}\right).
\end{equation}
\begin{equation}\label{eq:Cqrq}
C_q(r)^{1/(q-1)}=F_qr^{D_q}.
\end{equation}
Thus the generalised fractal dimension can be defined in the same way as the mass-radius fractal dimension, such that $D_q$ is given by
\begin{equation}\label{eq:DqDEF}
D_q=\frac{1}{q-1}\frac{d\log{C_q(r)}}{d\log{r}}.
\end{equation}
For some values of $q$ such that $q_i\neq q_j$ which satisfy $D_{q_i}=D_{q_j}$, i.e., $D_q$ is independent of $q$ and $r$, the distribution is called monofractal because its dimension is constant. In addition, if $D_q$ is equal to the Euclidean dimension, the distribution is homogeneous. For $q\geq1$, $D_q$ explores the scaling behavior in high-density environments within the distribution, which are associated mainly with clusters and superclusters, whereas for $q<1$, $D_q$ explores the behavior in low-density environments, i.e., those associated with \emph{voids} \citep{sarkar2009scale}. A full spectrum of the generalized fractal dimension provides detailed information about the entire distribution, whether in regions of high density or low density. This allows us to connect the concept of fractal dimension with statistical measures used to quantify the distribution of matter on large scales. If the distribution of galaxies undergoes a transition to homogeneity, all values of the fractal dimension tend to the dimension of the physical space, that is, $D_q\simeq D=3$ for any value of $q$; at small scales, we expect to see a spectrum of values of the fractal dimension all different from $3$, as strongly structures defined before of a homogeneity transition.

\section{Data samples}
\subsection{Observational holes and footprint}
BOSS is a SDSS-III project that mapped the spatial distribution of luminous red galaxies (LRGs) and quasars in two principal galaxy samples, LOWZ and CMASS, each one for the North Galactic Cap (NGC) and South Galactic Cap (SGC). BOSS uses optical fiber multi-object spectrograph with automated methods to controlling each fiber and effectively determining the regions of interest to record information from the target. The fibers can capture light from 1000 objects simultaneously, the objects were observed in a circular field of radius $1.49^\circ$ called \emph{tile} \citep{blanton2003efficient}. Although automated mechanisms were used, there are effects that cause erroneous or unmapped regions originating holes in the catalogues. Specifically, these effects are quantified with veto masks and are related with inherent process of observation producing sky regions that should be eliminated from the samples, including effects due to seeing, bright stars that saturate the detectors, trails caused by meteors and satellites or nearby objects, and failures in the position of the spectroscopic plates. Additionally there are regions that were masked where the imaging was unphotometric, usually due to too many blended objects in a single field or the image was identified as having critical problems in any photometric band \citep{Anderson11062014}.\\

The accumulation of these effects results in a distribution of observational holes in the galaxy samples. The effective area covered by the catalogue depends on the masks' quality and the sectors to be excluded from observation. The various types of masks depend exclusively on the catalogue. SDSS-III uses the following masks: \emph{Bright Star Mask}, \emph{Centerpost Mask}, \emph{Bad Field Mask}, and \emph{Collision Priority Mask} [for details on each mask see \citep{blanton2001efficient, smee2012multi, dawson2013baryon}]. All of these masks are exclusion masks and indicate the regions in which the data quality is unacceptable; i.e., if a point is within the mask, it will be excluded from the catalogue. According to \citet{blanton2001efficient}, these areas are relevant to and useful for large-scale structure studies. For fractal analysis \citet{yadav2010fractal} refers to some effects that may be caused by the geometry of the catalogues. Holes not only produce incompleteness regions, also modify the geometry of the masks by adding edges or borders, in this way \citet{Pan01032002} studied the boundary corrections in fractal analysis for the PSCz survey.\\

Although masks include observational holes, there is a footprint i.e., an observational template with the BOSS geometry that represents the ideal galaxy mapping without holes, and it covers the largest area possible. In this paper, we used the BOSS footprint \texttt{boss\_survey.ply}. The masks and BOSS footprint are designed in convex polygons that are independently generated for each of the 5 filters in the catalogue, in other terms is the union of weighted angular regions bounded by a number of edges. The geometry of the masks is given in terms of spherical polygons and arrays and the files are manipulated using the \emph{Mangle} code \citep{hamilton2004scheme}.\\

The CMASS sample has a mean completeness of 98.8\%, for the LOWZ sample it is 97.2\% \citep{reid2016sdss}. We use the NGC from the LOWZ sample in our analysis, this subset has 177,336 galaxies in 6,451 deg$^2$ and a veto area of 431 deg$^2$ producing a effective area of 5,836 deg$^2$ \citep{reid2016sdss}. The table \ref{tab:basic_props} lists some features about completeness for both the LOWZ and CMASS samples, like the total area, veto area, effective area and the total number of targets that were assigned a fiber within the survey footprint ($\bar{N}_{obs}$), the number of targets classified as stars ($\bar{N}_{star}$) or galaxies ($\bar{N}_{gal}$), and the number of targets for which the pipeline failed to find a robust classification and redshift ($\bar{N}_{fail})$.
\begin{table}
\begin{center}
\begin{tabular}{lrrrr}
Property & NGC & SGC &  NGC & SGC \\ \hline
Sample & \multicolumn{2}{c}{CMASS} & \multicolumn{2}{c}{LOWZ}   \\ \hline
$\bar{N}_{\rm gal}$ &607,357 & 228,990  & 177,336 & 132,191 \\
$\bar{N}_{\rm known}$ &11,449 & 1,841  & 140,444 & 13,073 \\
$\bar{N}_{\rm star}$ &14,556 & 8,262  & 1,043 & 976  \\
$\bar{N}_{\rm fail}$ &10,188 & 5,157  & 868 & 602 \\
$\bar{N}_{\rm cp}$ &34,151 & 11,163  & 4,459 & 4,422 \\
$\bar{N}_{\rm missed}$ &7,997 & 3,488  & 10,295 & 3,499 \\
$\bar{N}_{\rm used}$ &568,776 & 208,426  & 248,237 & 113,525\\
$\bar{N}_{\rm obs}$ &632,101 & 242,409  & 179,247 & 133,769 \\
$\bar{N}_{\rm targ}$ &685,698 & 258,901  & 334,445 & 154,763\\
Total area (deg$^2$) &7,429 & 2,823  & 6,451 & 2,823 \\
Veto area (deg$^2$) &495 & 263  & 431 & 264 \\
Used area (deg$^2$) &6,934 & 2,560 & 6,020 & 2,559  \\
Effective area (deg$^2$) &6,851 & 2,525 & 5,836 & 2,501  \\
Targets / deg$^2$ & 98.9& 101.1& 55.6& 60.5\\
\end{tabular}
\end{center}
\caption{Basic parameters of BOSS DR12 CMASS and LOWZ samples \citep{reid2016sdss}.}
\label{tab:basic_props}
\end{table}

\subsection{Synthetic samples}\label{subsec:Syntheticsamples}
\begin{figure}
	\centering
    \includegraphics[width=0.7\columnwidth]{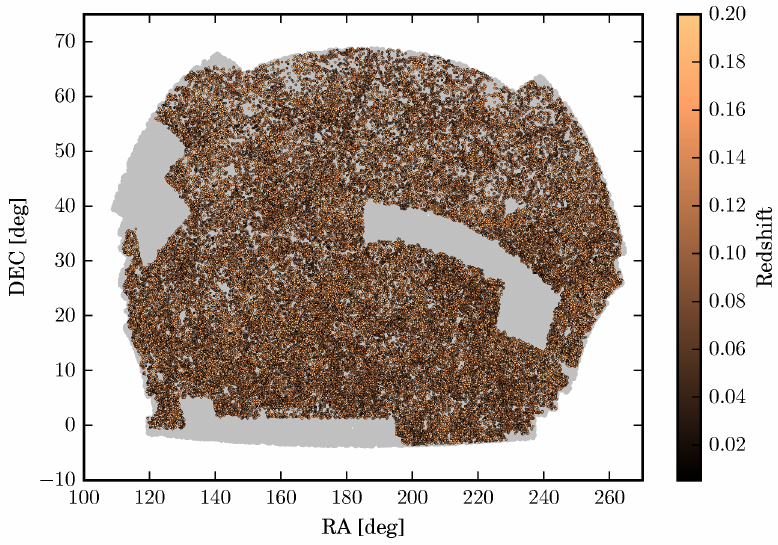}
    \caption{Galaxy sample in equatorial coordinates from LOWZ NGC catalogue. Colour bar indicates the redshift distribution limited to $0.002<z<0.2$ and gray areas indicates the BOSS footprint \texttt{boss\_survey.ply} upon completion of the survey.}
    \label{fig:HemisNorte} 
\end{figure}

Synthetic samples were built based on BOSS \emph{footprint} to cover the largest possible observation area in LOWZ NGC corresponding to 6,451 deg$^2$. The points forms a random distribution spatially uncorrelated that represents galaxies observed by BOSS in the same limits on redshift and equatorial coordinates right ascension and declination $(\alpha,~\delta)\equiv(RA,~DEC)$. BOSS footprint is made up of 19 polygons for the two galactic caps of the catalogue. In this paper we used only the NGC from the spectroscopic catalogue (see Fig.~\ref{fig:HemisNorte}). The southern hemisphere polygons were removed from \texttt{boss\_survey.ply}, then coordinates $(\alpha,~\delta)$ were assigned to each point within the \emph{footprint} using the Mangle code. First we selected a random polygon from the mask with a probability proportional to the product of the statistical weight and area of the polygon and a random point was generated within the circle. We checked whether the point was located within the polygon. If it was, the point is kept, otherwise the algorithm is repeated. The number density of points generated in the footprint corresponds to $74,959$, the same value as in the galaxy catalogue \texttt{galaxy\_DR12v5\_LOWZ\_North} in the BOSS Value-Added Catalogs available in \emph{SAS} \url{http://data.sdss3.org/sas/dr12/boss/lss/} (see Table~\ref{tab:samples}). 

\begin{table}
\begin{center}
\begin{tabular}{ccccc}
\hline
\multirow{2}{*}{\textbf{Sample}} & \multirow{2}{*}{\textbf{Points}} & \textbf{Area} & \textbf{Veto Area} & \textbf{RSPH}\\
              &              & (deg$^2$)       & (deg$^2$)            & (Mpc/h)        \\ \hline
Galaxy & 74,959 & 6,451,0 & 431 & \multirow{7}{*}{296.76} \\ \cline{ 1- 4}
0\% & 74,959 & 6,451.0 & 431 & \\ \cline{ 1- 4}
2\% & 73,412 & 6,322.0 & 560.02 &  \\ \cline{ 1- 4}
4\% & 71,886 & 6,193.0 & 689.04 & \\ \cline{ 1- 4}
6\% & 70,394 & 6,063.9 & 818.06 &  \\ \cline{ 1- 4}
8\% & 68,929 & 5,934.9 & 947.08 &  \\ \cline{ 1- 4}
10\% & 67,523 & 5,805.9 & 1,076.1 & \\ \hline
\end{tabular}
\end{center}
\caption{Basic properties of samples limited to $0.002<z<0.2$ used in the multifractal analysis. Percentages indicates the synthetic samples with a certain percentage of observational holes.}
\label{tab:samples}
\end{table}

Synthetic catalogues based on the BOSS footprint were limited in redshift to $0.002<z<0.2$, so the time interval between two events at these points is sufficiently small, and there are not significant changes in the large-scale structure patterns \citep{laporte2014evolution}. To model the redshift distribution of galaxies it is necessary to have a complete theoretical model that takes into account the behavior of the population of galaxies. Although a complete model is not available\citep{ross2012clustering}, the effects of the galaxy distribution in redshift can be known from information provided by the observed catalogues, some cosmological simulations, and random catalogues, considering that they always follow a radial selection function $n(z)$ \citep{ross2012clustering, kazin2010baryonic}. Naturally, the samples will be affected by the effects of the radial selection function, which result from the instrumental inability to detect faint galaxies at great distances. To determine $n(z)$, we used the shuffled method suggested by \citet{ross2012clustering} and \citet{kazin2010baryonic}, which consists of assigning randomly to each point of the sample a redshift value selected from the galaxy catalogue.\\

\begin{figure}
    \centering
	\includegraphics[width=0.6\linewidth,clip]{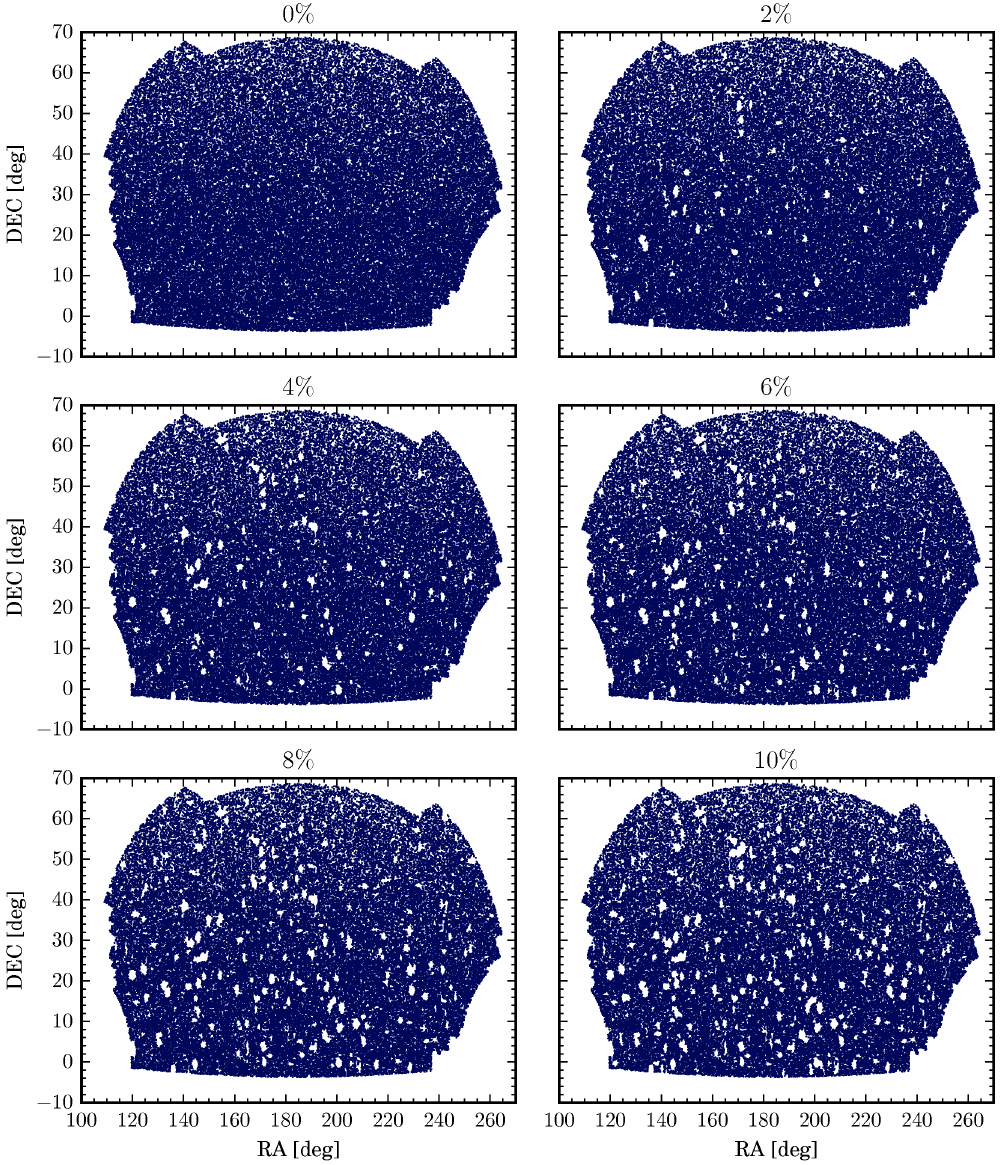}
    \caption{$\alpha-\delta$ projections of samples with holes limited to $0.002<z<0.2$. The percentages of holes were calculated on the basis of the total area from DR12 LOWZ NGC sample.}
	\label{fig:PorcentajeHuecos}
\end{figure}

To build the homogeneous synthetic samples the redshift values of DR12 were limited to the range $0.002<z<0.2$, next we used the shuffle method to distribute all points in the BOSS footprint, this is one of the least biased way to construct ramdom catalogues avoiding the excess of points in one coordinate \citep{ross2012clustering}. This method consists of assigning the redshift of a point of the random catalogue with the redshift of the observed data picked randomly, thus for each point in the synthetic samples $\alpha,~\delta$ and $z$ are known. The veto area in our sample from LOWZ NGC has a maximum size of 431 deg$^2$, this is approximately 7\% of the total observation area. These holes can be modeled as ellipses with semi-major and and semi-minor axis $a_{max}=b_{max}=3.00^\circ$. The footprint holes are generated randomly by assigning a random point with coordinates ($\alpha_{rnd},~\delta_{rnd}$); an ellipse centered at this point is drawn with semi-axes $a\in(0,3.00]^\circ$ and $b\in(0,3.00]^\circ$, and then the ellipse is removed from the footprint. This process is done for different percentages of holes based on the total number of points in the mask. Thus, we obtain the samples \emph{synthetic-0\%}, \emph{synthetic-2\%}, \emph{synthetic-4\%}, \emph{synthetic-6\%}, \emph{synthetic-8\%} and \emph{synthetic-10\%} for the corresponding percentages of holes. All samples are correlated because they are built cumulatively, i.e., \emph{synthetic-10\%} contains \emph{synthetic-8\%}, which contains \emph{synthetic-6\%} and so on. In Fig.~\ref{fig:PorcentajeHuecos} the projections in $\alpha-\delta$ are illustrated for the samples used in the multifractal analysis described in the next section. 

\section{Multifractal analysis}
\begin{figure}
\centering
\includegraphics[width=0.6\linewidth,clip]{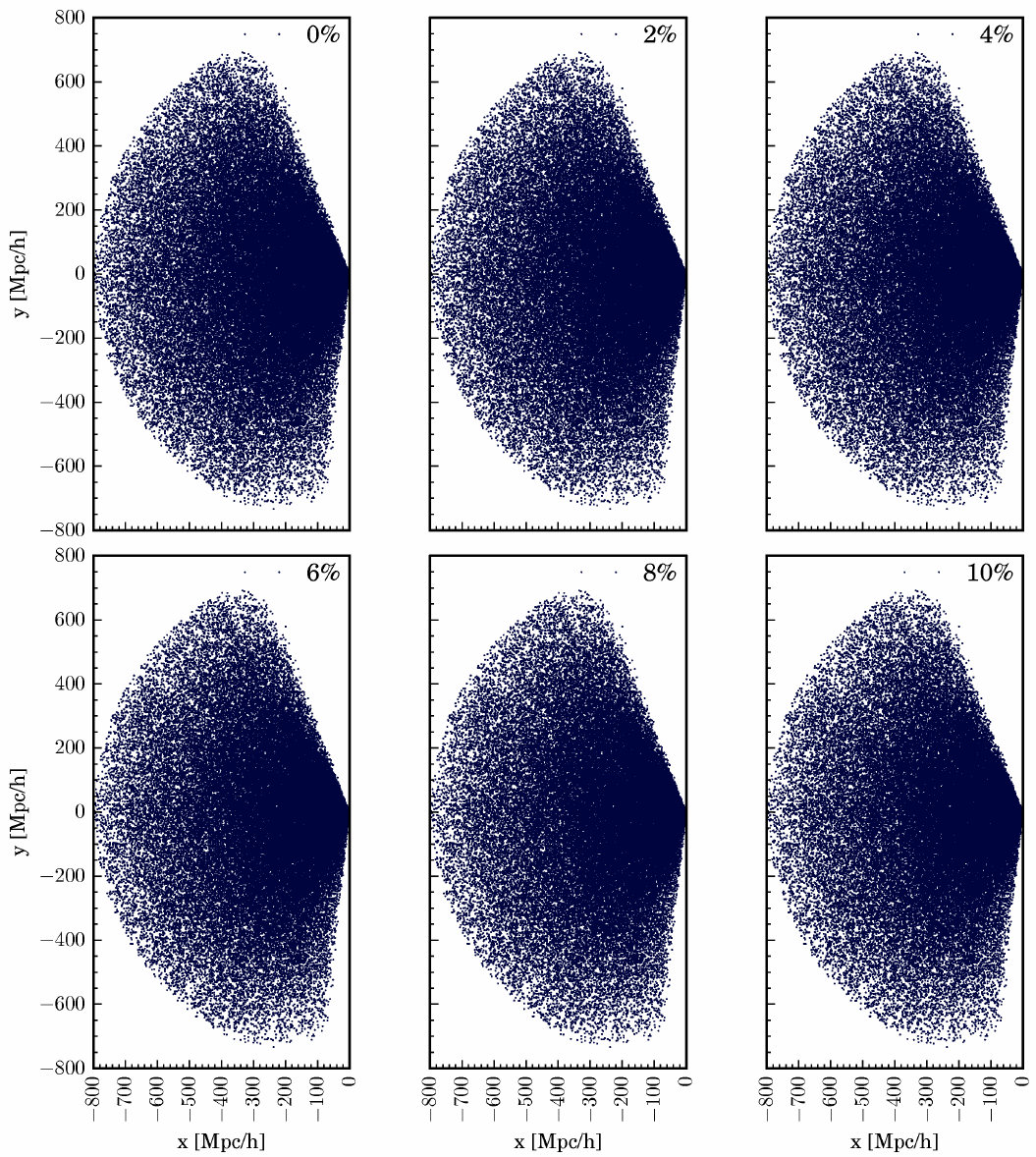}
\caption{Redshift-limited samples in comoving Cartesian coordinates for different hole percentage estimates on the BOSS footprint.}
\label{fig:MuestraEnComovil}
\end{figure}

The synthetic sample without holes is the densest sample with positions of $74,959$ objects in equatorial coordinates as shown in table~\ref{tab:samples}, the other samples have observational holes with values from 2\% to 10\% in the LOWZ NGC from BOSS footprint and contains a sampling of all points registered in the catalogue with a total area of $\sim6,451$ deg$^2$. In order to compare our results we also analyzed a real galaxy sample described in the previous section, this subset of the LOWZ (NGC) sample contains 74,959 galaxies limited in redshift to $0.002<z<0.2$ distributed in an effective area of 5,836 deg$^2$, equivalent to a synthetic sample with 6.7\% of holes respect to the footprint (see table \ref{tab:basic_props}). 
For each one of these samples the multifractal analysis was done in comoving Cartesian coordinates and the values of the parameters used in the cosmological distances calculations were $\Omega_m=0.308$, $\Omega_k=0.0$, $\Omega_\Lambda=0.691$, $h=0.73$, $\omega=-0.980$ according to Planck 2015 Results \citep{planck2015planck}. To perform the analysis we follow \citep{joyce2005basic} and we made a sampling within the limited redshift samples taking in account the radius of a larger sphere than these can contain. According to \citet{gabrielli2006statistical} the effective depth $R_s$ for all samples given its limits in $\alpha-\delta$ can be determined by equation~(\ref{eq:RsFractal}).	

\begin{equation}\label{eq:RsFractal}
R_s=\frac{R_d\sin(\delta\theta/2)}{1+\sin(\delta\theta/2)},
\end{equation}

where $R_d$ corresponds to the maximum radial distance from the sample, and $\delta\theta=\text{min}(\alpha_2-\alpha_1,~\delta_2-\delta_1)$, where $\alpha_2,~\alpha_1$ are the limits in right ascension, and $\delta_2,~\delta_1$ are the limits in  declination. In Fig.~\ref{fig:MuestraEnComovil}, a projection in comoving Cartesian coordinates is shown for different samples used in our analysis. The generalised correlation integral $C_q$ and the fractal dimension $D_q$ were calculated for each synthetic sample. The statistical uncertainty of $C_q$ was determined using the equation~(\ref{eq:Cq}) for the corresponds error propagation.

\begin{equation}\label{eq:PropErrorCq}
\Delta C_q(r)=\frac{1}{MN}\sum_{i=1}^M(q-1)\left[n_i(r)\right]^{q-2}\Delta n_i(r).
\end{equation}

\begin{figure*}
\centering
\includegraphics[width=1.0\linewidth,clip]{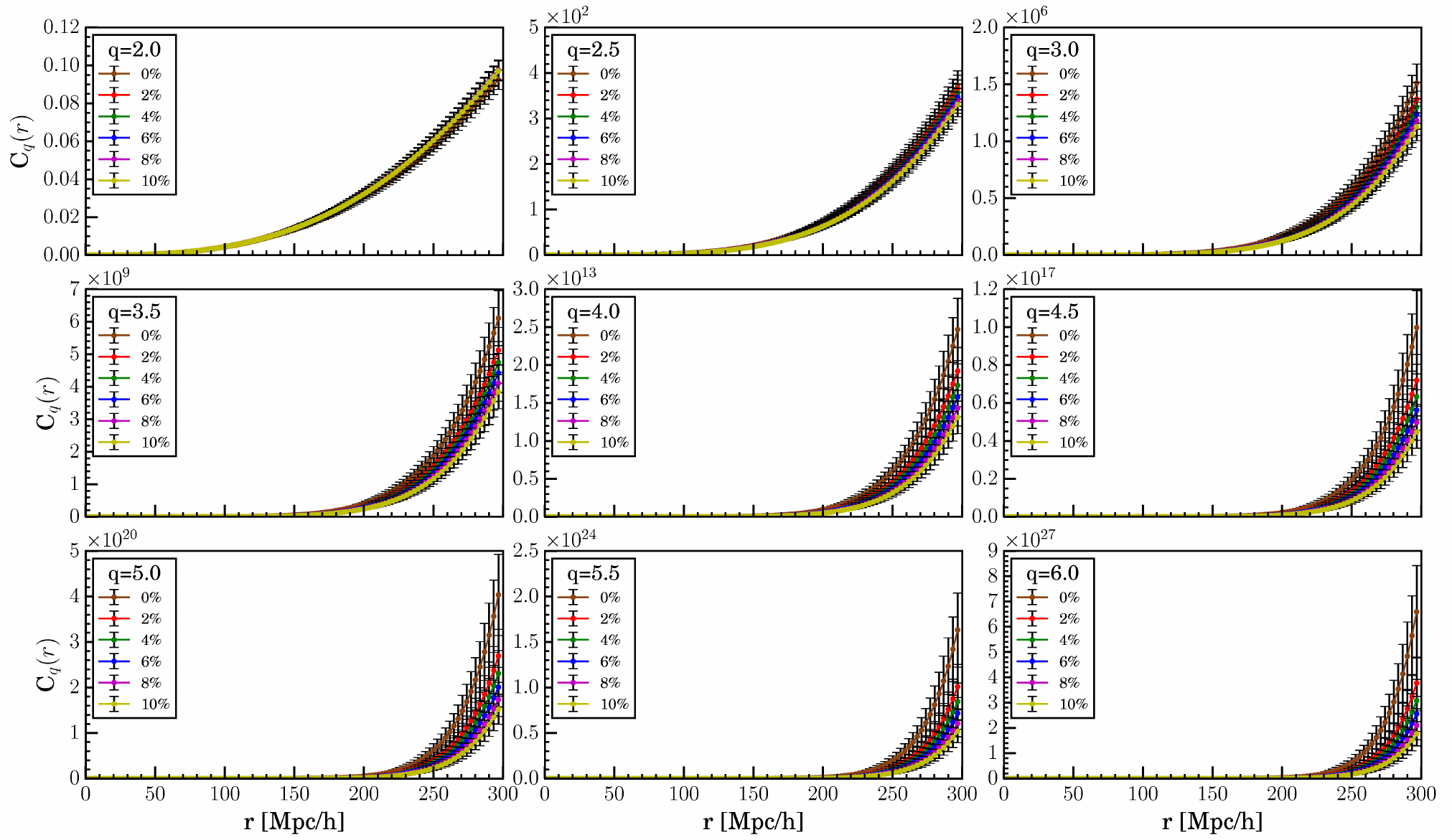}
\caption{Generalized correlation integral $C_q(r)$ for all $q$ values studied in the synthetic samples.}
\label{fig:Cq_BOSS_Huecos}
\end{figure*}

The uncertainty $\Delta n_i(r)$ is determined numerically counting the number of points that could have been included in $n_i(r)$ but were omitted because given the uncertainty in the position, that is, those points that statistically are outside the spherical shell despite being very close to its borders \citep{chacon2012millennium}. The step distance in the $C_q$ calculations is $\Delta r=r_{i}-r_{i-1}=1.0~Mpc/h$, this value corresponds to the average distance between two galaxies in clusters \citep{narlikar2002introduction, sharan2009spacetime, martinez2010statistics}. The $C_q$ value was determined for each sample using 9 structure parameters in the range $2\leq q\leq6$ in order to compare the results with those of \citep{chaconstudy, scrimgeour2012wigglez, PanColes2000}. Fig.~\ref{fig:Cq_BOSS_Huecos} shows the behavior of $C_q$ with respect to comoving distance, where all values $q\geq0$ indicates high-density regions within the distribution. This plot is consistent with the results obtained by \citet{chaconstudy}, \citet{PanColes2000} and \citet{sarkar2009scale}, where the set of integrals shows how the average number of neighbours change around the centers, they increase for $q\geq2$, and these functions indicate the probability to get a pair of points in a distance less than or equal to $r$. All samples studied exhibit the same behavior for $C_q$ independent of the hole density, thus the fractal dimension is convergent and it is similar for different values of $q$.

\begin{figure}
\centering
\includegraphics[width=.7\linewidth,clip]{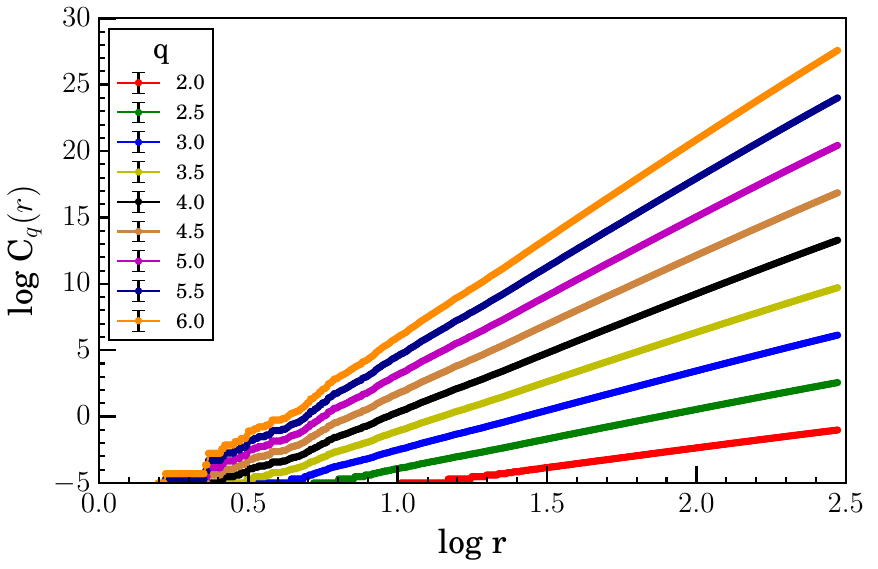}
\caption{$\log{C_q(r)}$ versus $\log{r}$ for all synthetic samples and each structure parameter $q\in[2,~6]$. Each color line contains all synthetic samples given a specific value of $q$.}
\label{fig:LogCq-Logr}
\end{figure}

The fractal dimension $D_q(r)$ and its uncertainty were calculated from the generalized correlation integral (see equation~(\ref{eq:DqDEF})) and its corresponding error propagation. For radial depths that reach as large as 296 Mpc/h the sliding window technique presented by \citet{martinez2003statistics} and \citet{rodrigues2004self} was applied to the logarithm of the correlation integral as a function of the logarithm of the comoving distance $r$. In this case we have a linear relationship \citep{pietronero2004statistical, chacon2012millennium}, i.e., $\log(C_q)\propto\log(r)$, where the constant of proportionality is related to the fractal dimension, see equation~(\ref{eq:ExpanLOG}). By successive approximations to a line segments the slope $\tau(q)=\Delta\log(C_q)/\Delta\log(r)$ allows calculate the generalized fractal dimension as $D_q=\frac{1}{q-1}\tau(q)$ and the intersection $\frac{1}{q-1}\log(F_q)$ with their corresponding statistical uncertainties. This method was validated using a main sample without holes which represents an ideal sample based on the BOSS footprint with 3,273,548 points distributed uniformly across the mask limited to $0.002<z<0.2$, as expected in this sample fractal behavior does not occur. For all values of the parameter structure the fractal dimension $D_q$ converges rapidly to the dimension of physical space, $D=3$. The behavior of $\log{C_q(r)}$ versus $\log{r}$ for all synthetic samples can be clearly seen in Fig.~\ref{fig:LogCq-Logr}. Each structure parameter $q\in[2,~6]$ and each color line contains all synthetic samples (percentages of holes from 0\% to 10\%), this means that all samples with holes less or equal than 10\% overlap each other and they have the same behavior independent of the $q$ value. From these plots it is possible to obtain the multifractal spectrum of $D_q(r)$ as a function of the comoving distance $r$ for all samples.\\

\begin{figure}
\centering
\includegraphics[width=.7\linewidth,clip]{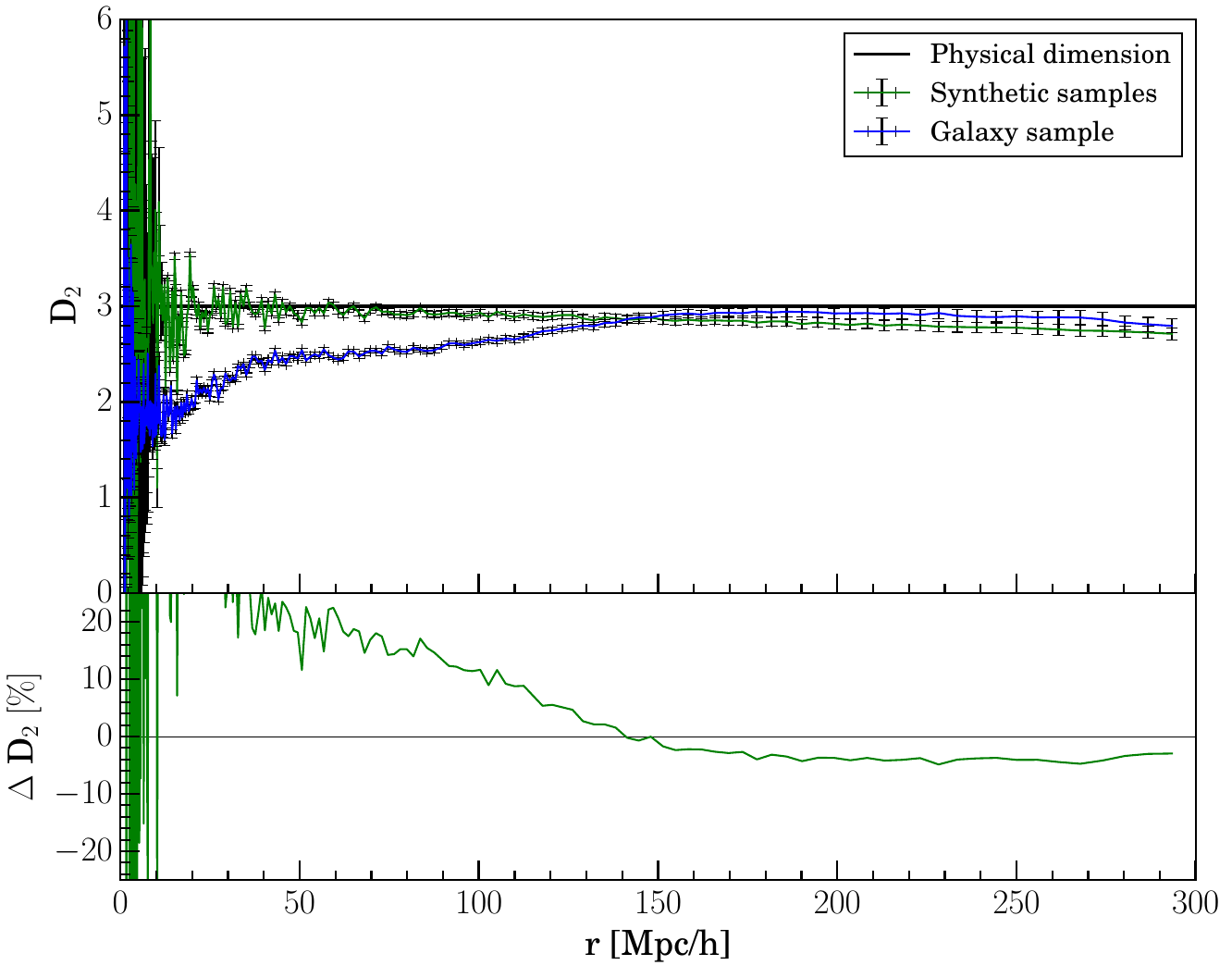}
\caption{Upper panel: Multifractal spectrum of $D_2(r)$ (structure parameter $q=2.0$), as a function of the comoving distance $r$ for all synthetic samples (green line) and the galaxy sample (blue line). Lower panel: percentage difference in fractal dimension between synthetic and galaxy samples where $\Delta D_2=100(D_{2_{synth}}-D_{2_{gal}})/D_{2_{gal}}$.}
\label{fig:Dq_VS_r}
\end{figure}

An especially interesting case is $q=2$ because the correlation dimension $D_2$ is related to the homogeneity scale $r_H$ and the usual two-point correlation function for a sample displaying cosmic homogeneity \citep{peebles1980large,peebles1989fractal,PanColes2000}. Fig.~\ref{fig:Dq_VS_r} shows the spectrum of fractal dimension $D_q$ and its dependency with radial distance for synthetic samples. For $q=2$ the dimension tends to $D=3$ as $r$ increases, at scales below 50Mpc/h the distribution it is highly noisy indicating that it is grouped into small spatial regions. In high-density regions there is a strong tendency to homogeneity because the values of the fractal dimension are very close to the physical space dimension. $D_{q\geq2}$ increase at large $r$ values to reach homogeneity in concordance with other analysis, this means that on average the space is statistically filled at depths greater than $r_H$.\\

This can be better appreciated in the lower panel of Fig.~\ref{fig:Dq_VS_r} that shows the percentage difference of fractal dimension between synthetic and galaxy samples $\Delta D_2=100(D_{2_{synth}}-D_{2_{gal}})/D_{2_{gal}}$. On scales below 140~Mpc/h synthetic samples exhibit a fractal dimension significantly different respect to the galaxy sample with deviations up to 10\% independent of the hole percentages. On scales above 150~Mpc/h the dimension stabilizes at $D_2=2.82$ and it is compatible with the galaxy sample with a relative difference of $3.5\%$.\\

\begin{figure}
\centering
\includegraphics[width=0.7\linewidth,clip]{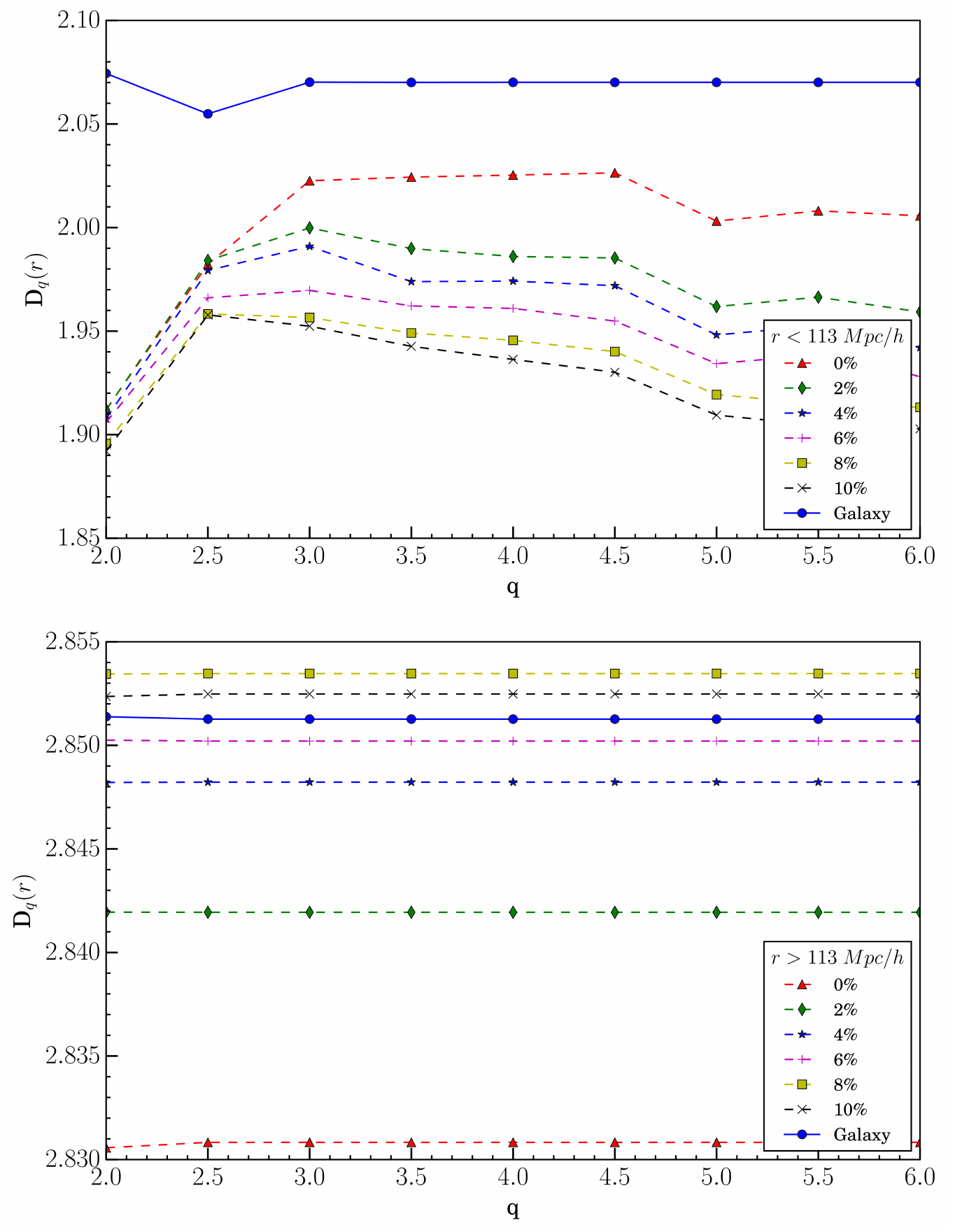}
\caption{Comparison of fractal dimension at $r_H$ as a function of the structure parameter $q$ for each sample. Top: average dimension for distances $r<113~Mp/h$, bottom: average dimension for distances $r\geq113~Mp/h$.}
\label{fig:Dq_VS_q-CadaR-Huecos}
\end{figure}

The behavior of $D_q$ for low and over density regions in terms of the structure parameter in the range $2\leq q\leq6$ is shown in Fig.~\ref{fig:Dq_VS_q-CadaR-Huecos}. For percentages of holes between $0-10\%$ and $r<113~Mpc/h$ the dimension changes with $1.89^{\pm0.09}\leq D_q\leq2.08^{\pm0.09}$ exhibiting a fractal behavior, and for the same percentages at large scales $r>113~Mpc/h$ the fractal behavior disappears and the dimension is stable between $2.83^{\pm0.09}\leq D_q\leq2.86^{\pm0.09}$. Following \citet{scrimgeour2012wigglez}, we define an homogeneity scale $r_H$ of the correlation dimension as the scale for which the value of $D_2(r)$ is within 1\% of $D_2=3$, i.e. $D_2(r=r_H)=2.97$. This definition of $r_H$ does not depend on the catalogue and is not affected by the sample size, also it can be used to compare different measurements as long as the same definitions are used in all cases \citep{scrimgeour2012wigglez}, other authors who have also used this same definition for $r_H$ are \citep{PierrosNtelis2017} and \citep{Li2015A}.\\

\begin{figure}
\centering
\includegraphics[width=.7\linewidth,clip]{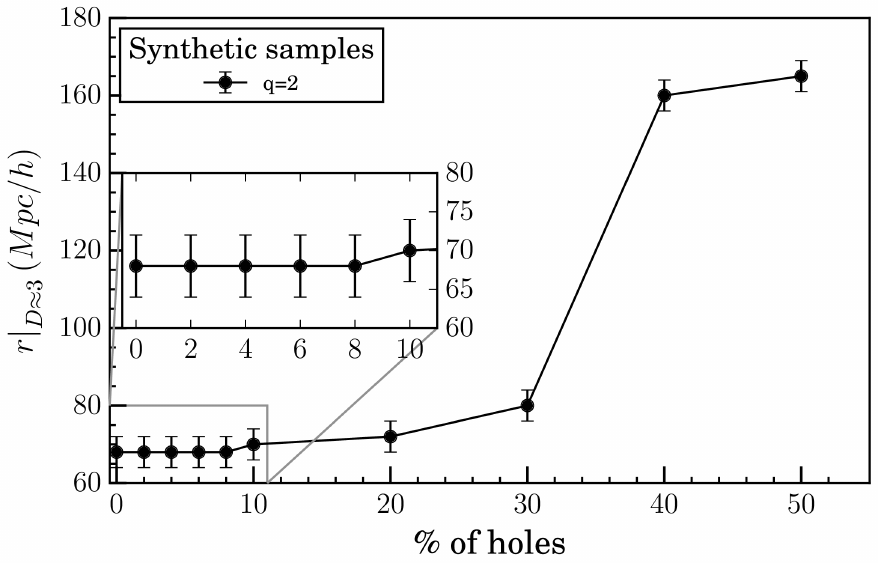}
\caption{Homogeneity scale $r_H$ as a function of the hole density for all synthetic samples.}
\label{fig:PorcenHuecos_VS_Escala_Homoge}
\end{figure}

Using this definition we calculate the homogeneity scale of each sample. The dependence of $r_H$ as a function of the hole density for synthetic samples is presented in Fig.~\ref{fig:PorcenHuecos_VS_Escala_Homoge}. This plot shows the correlation for samples with hole density below 10\%, in which case $r_H$ is nearly constant evidencing a statistical homogeneity at this scale $83\pm1$~Mpc/h, besides, to illustrate the behavior with large hole densities we completed our analysis with four synthetic samples corresponding to 20\%, 30\%, 40\%, 50\% of holes each one, built following the same procedure mentioned in subsection \ref{subsec:Syntheticsamples}. Even more, we found a correlation between the hole density of the masks and the homogeneity scale, particularly Fig. \ref{fig:PorcenHuecos_VS_Escala_Homoge} shows the evolution of the homogeneity scale for $D_2$ as a function of the hole density. When observational holes are over 10\% its cause shifts in the homogeneity scale and for veto percentages near to 6\% the dimensions are consistent with the galaxy sample with deviations less than 1\% around $D_q=2.85$ at $r>113~Mpc/h$.

\section{Conclusions}
We analyzed the multifractal behavior in samples built from the BOSS footprint and investigate if these holes modify the scale of cosmic homogeneity. We include a random distribution of observational holes related with the veto masks of LOWZ NGC sample and it were compared with a real galaxy sample with same geometrical properties (angular distribution and radial selection function) but with a different point distribution. The results we have presented here provide evidence that the effect of veto areas can cause shift in homogeneity scales.\\

According to our results the spectrum of fractal dimension gives enough information to show a transition of the homogeneity scale at a specific scale. This justifies the use of statistically homogeneous samples following the observed radial selection function and geometric features of the galaxy masks. Our calculations take into account the maximum radius of the comoving circumscribing sphere of the edges for each sample, therefore the fractal study can be performed by small and successive iterations within this sphere. The results we have presented in this paper provided strong evidence that the hole distribution it is an important parameter to consider in the calculus of the homogeneity scale using fractal and multifractal analysis. In particular we showed that in synthetic samples the generalized dimension varies with the scale, at small scales there are a fractal behavior for all hole percentages studied, this behavior disappears at large scales until reaching the statistical cosmic homogeneity at $D_q\approx3$ for scales close to 113 Mpc/h in concordance with other analysis.\\

For all samples studied with holes between $0-10\%$ and for each $q\geq2$, the values of the scale at the homogeneity transition are around $r>113~Mpc/h$. The dimension is stable between $2.83^{\pm0.09}\leq D_q\leq2.86^{\pm0.09}$ evidencing a statistical homogeneity at this scale with a deviation of 3\% around $D=3$. This means that the distribution,
both synthetic and galaxies samples, has a tendency to fill the physical space on larger scales and its supports the standard cosmological framework. For veto percentages near to 6\% the dimensions are consistent with the galaxy sample with deviations less than 1\% around $D_q=2.85$ at $r>113~Mpc/h$.\\

Deviations of exact value $D=3$ can be explained because holes induce small fluctuations around the homogeneity scale, according to Fig. \ref{fig:PorcenHuecos_VS_Escala_Homoge} $r_H$ is shifted to bigger values when the hole distribution is bigger too, hence the set dimension before this scale $r_H$ is less than $D=3$. This behavior depends exclusively on the holes in the sample and suggest that veto areas in the masks must be taken into account in analysis of homogeneity scale measured from galaxy samples. Consequently future results about homogeneity scale based in fractal analyses must be corrected by observational holes and regions of incompleteness in the geometry of the galaxy catalogue if the size of the veto mask it is significant. Our future work on this problem is to examine this effect in different galaxy samples at high redshift, study the case of $q\leq0$ and examine lacunarity spectrum to quantify the hole distribution.

\section*{Acknowledgements}
JEGF acknowledges financial support from the Facultad de Ciencias at Universidad Nacional de Colombia - Sede Bogot\'a to perform master studies. The authors thank Dr. C\'esar Alexander Chac\'on for providing the code for calculating fractal quantities and Dr. Florian Beutler for suggestions about BOSS footprint.\\
Funding for SDSS-III has been provided by the Alfred P. Sloan Foundation, the Participating Institutions, the National Science Foundation, and the U.S. Department of Energy Office of Science. The SDSS-III web site is http://www.sdss3.org/.\\
SDSS-III is managed by the Astrophysical Research Consortium for the Participating Institutions of the SDSS-III Collaboration including the University of Arizona, the Brazilian Participation Group, Brookhaven National Laboratory, Carnegie Mellon University, University of Florida, the French Participation Group, the German Participation Group, Harvard University, the Instituto de Astrofisica de Canarias, the Michigan State/Notre Dame/JINA Participation Group, Johns Hopkins University, Lawrence Berkeley National Laboratory, Max Planck Institute for Astrophysics, Max Planck Institute for Extraterrestrial Physics, New Mexico State University, New York University, Ohio State University, Pennsylvania State University, University of Portsmouth, Princeton University, the Spanish Participation Group, University of Tokyo, University of Utah, Vanderbilt University, University of Virginia, University of Washington, and Yale University.

\section*{References}
\bibliography{references}
\bibliographystyle{apalike}

\end{document}